\documentclass[12pt]{iopart}
\usepackage{epsfig,mathrsfs,graphicx,iopams}

\begin{document}
\title{ Quantum Objective Realism }
\author{Adam Bednorz}
\address{Faculty of Physics, University of Warsaw, ul. Pasteura 5, PL02-093 Warsaw, Poland}
\ead{Adam.Bednorz@fuw.edu.pl}

\begin{abstract}
The question whether quantum measurements reflect some underlying objective reality has no generally accepted answer.
We show that description of such reality is possible under natural conditions such as linearity and causality, although in terms of moments and cumulants of finite order and without relativistic invariance. The proposed construction of observations' probability distribution originates from weak, noninvasive measurements, with detection error replaced by some external finite noise. The noise allows to construct microscopic objective reality, but remains dynamically decoupled and hence unobservable at the macroscopic level.

\end{abstract}

\submitto{\NJP}
\maketitle

\section{Introduction}
The question whether quantum states -- or at least their observable properties -- are real, independent from deliberate measurements, has been asked already in the early days of quantum mechanics, leading to notorious Einstein-Podolsky-Rosen paradox \cite{epr}. At the same time a realistic interpretation appeared in the form of Bohm-de Broglie pilot wave theory \cite{bohm}, which by construction is impossible to falsify. However, it never gained wide-spread acceptance due to its manifest nonlinearity,  difficulties with generalizations and nonlocality. For this last reason,  this interpretation also fails to construct \emph{local} reality. 
As shown later by Bell \cite{bell}, local reality is in conflict with the results of a simple ideal experiment (Bell test), never conclusively verified \cite{belx}  due to experimental nonidealities/loopholes \cite{loop}. Pusey, Rudolph and Barrett (PBR)\cite{pbr} and Colbeck and Renner (CR) \cite{colren} have recently shown that local reality cannot hold under some assumptions of (almost) perfect measurements.

A different viewpoint on quantum reality assumes that the measurement or more precisely collapse takes place 
spontaneously and continuously, independently of our will, hence objectively, blurred by random noise. In this case objective reality does not necessarily applies to the formal quantum state itself but rather its observations.  Such continuous objective collapse theories invented by Ghirardi, Rimini and Weber (GRW)
\cite{grw,csl}  are much more attractive to become candidates for objective realism than Bohm-de Broglie because they are constructed using standard theory of quantum measurements, projection postulate
\cite{neumann}, but generalized  to positive operator-valued measures (POVM) \cite{povm}.
Strong measurements (projections), general POVMs and objective collapse could in principle reveal reality,
which is the major strength of GRW approach.
However, the two problems with collapse -- absent in Bohm-de Brogile -- are that it (i) makes the quantum dynamics nonunitary (interrupted by irreversibility of collapse) and (ii) leads to constant heating of the system which is the whole universe, although at very small  rate. The second problem can be resolved introducing a heat sink, then it leads to thermalization of the universe.

Here we construct quantum objective realism for observations of a quantum state that on one hand leaves the quantum dynamics unitary (dynamically reversible) without heating, being linear and causal on the other hand. The question of locality requires introducing freedom of choice, which will be the future step.
The construction is analogous to weak \emph{limit} of continuous collapse by GRW \cite{grw,csl} or weak measurement \cite{weak}. Weak measurement is in fact noninvasive as it does not disturb the measured system, but only in the zero-strength limit. Such measurements are beyond PBR and CR assumptions, requiring rather strong than weak measurements. The problem with weak, noninvasive measurements is that direct interpretation of their result leads to strange \emph{weak values}, which may exceed intuitive spectrum (being e.g. $100$ for the spin $1/2$ \cite{weak}). This is possible because weak measurements make a natural separation between (large, or infinite in the weak limit) detection noise/error, independent of the measured system, and system-only-dependent statistical measure, which reveals to be a quasiprobability rather than a standard nonnegative probability \cite{bednorz:prl10,abz}. On one hand the limiting quasiprobability alone cannot  describe objective realism as it can be negative, on the other hand the detection noise cannot be made infinite, either. 

We propose to take the limiting quasiprobability and add a finite amount of external (replacing infinite detection) noise, only to make it positive. It can be interpreted as GRW collapse without disturbance. This applies to both finite and continuous but countable variables (e.g. regularized charge density),
although for an infinite number of variables only moments and cumulants up to a finite order can strictly correspond to positive probability.
We will show that this is always possible within the realistic constraints, that the noise should be negligible and unobservable on macroscopic level. However, it is impossible to construct objective realism, combined with unitary dynamics, linearity and causality, which is invariant with respect to relativistic frame transformations. Although the underlying quasiprobability can be made invariant, the added noise cannot. Moreover, time symmetry must be also broken \cite{bfb}. 

The paper is organized as follows. We first construct objective probability by a convolution of noise and quantum quasiprobability. Then the general procedure of adding noise to get positive probability from quasiprobability is outlined. Next, the problem of broken relativistic invariance is discussed and finally realistic constraints of the noise are proposed. Some mathematical details are moved to appendices.

\section{Construction}
\label{secon}

We shall use standard relativistic notation, with $c=\hbar=k_B=1$, real coordinates $x=(x^\mu, \mu=0,1,2,3)$ with  single time $x^0$, (flat) metric tensor $g^{\mu\nu}=g_{\mu\nu}=\mathrm{diag}(1,-1,-1,-1)$ and Minkowski product $x\cdot y= x^\mu y_\mu=\sum_\mu x^\mu y_\mu$, shift $x_\mu=g_{\mu\nu}x^\nu$, derivatives $\partial_\mu=\partial/\partial x^\mu$, integration measure $\rmd x=\rmd x^0\rmd x^1\rmd x^2\rmd x^3$ and $\delta(x-y)=\prod_\mu \delta(x^\mu-y^\mu)$. We shall not treat time and position as physical quantities, only descriptors of spacetime.
We assume that observable physical quantities $a$ in spacetime such as field  $\phi(x)$ and current $j^\mu(x)$ are \emph{objectively real} so they always have some values. However, they can be random, with some non-negative and normalized probability distribution functional $P[\phi,j,...]$.
We also assume that $P$ is a convolution of some external (non-quantum) noise probability $N$ and a quantum quasiprobability $Q$,
\begin{equation}
P[a]=N\ast Q=\int \mathrm{D}a'  N[a']Q(a-a').\label{conv}
\end{equation}
In the case when $Q$ is a normal (positive) probability such a convolution appears naturally in classical physics when two independent probability distributions are combined.

Now we have to make connection with quantum mechanics.
Each observable $\hat{A}_j$ (the index will be dropped when unimportant) can have its counterpart in objective reality $a_j(x)$. Some selection of observables is necessary, we can take at least those appearing in the macroscopic world, such as charge, current, electromagnetic field. Instead of measurements, which are intuitively invasive and require deliberate actions, we shall describe observations, which are passive and nondisturbing. Observations take place in the whole spacetime so this description is in practice informationally complete even without explicit reconstruction of the full quantum state.
From the property of convolution we have adding up averages and correlations,
$\langle a\rangle_P=\langle a\rangle_N+\langle a\rangle_Q$ and $\langle\delta a\delta b\rangle_P=
\langle\delta a\delta b\rangle_N+\langle\delta a\delta b\rangle_Q$ for $\delta a=a-\langle a\rangle$.
Since $N$ is simply adding its own contribution, we will \emph{postulate} that objective correlators of observations like $\langle a_1(x_1)\cdots a_n(x_n)\rangle_Q$ should be related to their equivalents in quantum mechanics, which will be represented by $Q$
replacing $a(x)$ in the correlator by  a superoperator $\int \rmd x' \check{A}^{x-x'}(x')$ in Heisenberg picture and perform time order \cite{bbrb}:
\begin{equation}
\langle a_1(x_1)\cdots a_n(x_n)\rangle_Q=\int \mathrm{d}^nx'\:\mathcal T\langle\check{A}_n^{x_n-x'_n}(x'_n)
\cdots\check{A}_1^{x_1-x'_1}(x'_1)\rangle.
\label{avgabs}
\end{equation}
Here, $\mathcal T$ denotes time order with respect to the arguments $x^{\prime 0}$ in
brackets, $\langle\cdots\rangle$ means quantum average $\mathrm{Tr}\cdots\hat{\rho}$ with the initial state density matrix $\hat\rho$. 
Superoperators should fulfill linearity and causality -- outcomes can depend only linearly on operators within their lightcones, which imposes the form
\begin{equation}
\check{A}^{x-x'}(x')
=g(x-x')\check{A}^c(x')+f(x-x')\check{A}^q(x')/2\:.\label{avg2}
\end{equation}
The superoperators $\check{A}^{c/q}$ \cite{zwanzig} act on any
operator $\hat{X}$ as an anticommutator/commutator:
$\check{A}^c\hat{X}=\{\hat{A},\hat{X}\}/2$ and
$\check{A}^q\hat{X}=[\hat{A},\hat{X}]/i$.  Alternatively $2\check{A}^c=\check{A}^++\check{A}^-$ and $i\check{A}^q=\check{A}^+-\check{A}^-$ with $\check{A}^+\hat{X}=\hat{A}\hat{X}$ and $\check{A}^-\hat{X}=\hat{X}\hat{A}$.
In the above expressions
we assumed that observations are stationary so that only relative differences $x-x'$ matter.
Please note that the time order is irrelevant for spacelike intervals ($y=x-x'$ with $y\cdot y<0$) because then the operators commute.
The commuting property can be shown using closed time path theory \cite{ctp,ab13} for each Feynman diagram \cite{peskin}. Hence, time order is here in fact causal order, consistent with relativity.
The quasiprobability $Q$ itself can be formally expressed by Dirac $\delta$ with operators inside,
\begin{equation}
Q=\int \mathrm{d}^nx'\:\mathcal T\langle\delta(a_n-\check{A}_n^{x_n-x'_n}(x'_n))
\cdots\delta(a_1-\check{A}_1^{x_1-x'_1}(x'_1))\rangle.
\label{qqq}
\end{equation}
Note that time ordering $\mathcal T$ makes it important to expand $\delta$ as a time-ordered series if $x'$ may overlap.

We will also assume that the average of single observation coincides with the usual average for projective measurements, i.e. $\langle a(x)\rangle=\langle\hat{A}(x)\rangle$. 
This implies $g(x-x')=\delta(x-x')$. Other choices of $g$ simply mimic the effect of classical filters. 
However, its is reasonable to expect \emph{causality}, namely $g(x)=0$ for $x^0<0$, which implies zero also for $x\cdot x\leq 0$ by invariance.
Thus the only freedom left is the
choice of the real memory function $f$ that multiplies $\check{A}^q$. Note that $f(x)$
can be non-zero for $x^0>0$ without violating causality, since it is
accompanied by $\check{A}^q$ and only future observations are
affected. For the last observations, future effects disappear because
the leftmost $\check{A}^q$ vanishes under the trace in Eq. (\ref{avgabs}).
In the Markovian case $f=0$.  The formula (\ref{avgabs}) can be derived from the weak measurement approach, see \ref{appa}. Note  however that the above construction is beyond standard theory of noninvasive measurements \cite{weak}, where either the noise $N$ was infinite, or $Q$ contained the effect of detector's disturbance.
However, $N$ is decoupled from quantum dynamics, so no detection process is able to amplify it to macroscopic level.
In fact $N$ allows to assign objective reality to microscopic systems, but remains negligible, and hence unobservable, macroscopically. 

\section{Quasiprobability}
\label{secq}

 In the case of position $\hat{x}$ and momentum $\hat{p}$ for $\hat{A}_1$ and $\hat{A}_2$ (in arbitrary order) with $f=0$,
 $Q$ becomes well-known Wigner function, which is a quasiprobability \cite{wigner}, but can be made positive adding noise and obtaining positive Husimi function \cite{hus}. For $f=1/\pi t$ and harmonic system\cite{bbrb} one reveals another quasiprobability --
 Glauber-Sudarshan function \cite{glau}.
The definition (\ref{qqq}) is a generalization of Wigner function to arbitrary set of time-dependent observables.
Let us demonstrate the nontrivial effect of the noise $N$ on the quantum statistics in a simple two-level model, with states 
$|\pm\rangle$, and observables (Pauli matrices) $\hat{A}=|+\rangle\langle -|+|-\rangle\langle +|$ and 
$\hat{B}=|+\rangle\langle +|-|-\rangle\langle -|$ analogously to \cite{weak}. The observables correspond to observations $a$, $b$, respectively. We have to specify time order of the observations. For simplicity we take $f=0$ (Markovian case). Let us prepare the state $|+\rangle\langle +|$ and ask for the joint probability $a$ and $b$, with first $\hat{A}$ then $\hat{B}$. The quasiprobability $Q$ in this case is not continuous but a kind of discrete Wigner function (consisting of $\delta$ peaks in continuous representation), and can be calculated exactly, with the results in Table \ref{tab1}, see
also detailed calculation in \ref{appx}.
\begin{table}
\begin{tabular}{|r|r|r|r|r|}
\hline
$Q(a,b)$&$-1$&$0$&$+1$&$a$\\
\hline
$-1$&$1/4$&$\mathbf{-1/2}$&$1/4$&\\
\hline
$+1$&$1/4$&$+1/2$&$1/4$&\\
\hline
$b$&&&&\\
\hline
\end{tabular}
\caption{Quasiprobability as a function of $a$ and $b$ of the observables first $A$ then $B$}\label{tab1}
\end{table}

The negative quasiprobability appears in the entry $Q(a=0,b=-1)=-1/2$.
The probability $P$ can be written in terms of the noise $N$
\begin{eqnarray}
&&4P(a,b)=N(a-1,b-1)+2N(a,b-1)+N(a+1,b-1)\nonumber\\
&&+N(a-1,b+1)-2N(a,b+1)+N(a+1,b+1)\label{pnn}
\end{eqnarray}
 Assuming a Gaussian noise $N(a,b)\propto \rme^{-\alpha a^2-\beta b^2}$ (with $\alpha,\beta>0$) helps to make it  positive but not everywhere, because 
$P(0,b)\propto \rme^{\beta(4 b-1)}(1+\rme^{\alpha/2}\sinh\alpha)-\rme^{\alpha/2}\sinh\alpha$ which is negative for $b\to  -\infty$.
The problem is that Gaussian decays too rapidly. However, taking $N\propto \rme^{-\alpha|a|-\beta|b|}$
with sufficiently small $\alpha,\beta$ we get everywhere positive $P$. In fact such an exponential noise gives positive $P$ for any \emph{finite} number of observables in a finite space. Unfortunately this is not the case of infinite sequences, similarly as in PBR and CR arguments. Suppose that we demand objective reality for any sequence containing $\hat{A}$ and $\hat{B}$. For instance, we can map the problem on values for  time resolved observable $\hat{B}(x^0)$ by introducing the Hamiltonian $\hat{H}=\omega i(|-\rangle\langle +|-|+\rangle\langle -|)$ which periodically alternates between $A$ and $B$. Let us take an arbitrarily large sequence $\hat{A}\hat{B}\hat{B}\hat{B}\hat{B}....$ Then all last $\hat{B}$s will have the same value for the quasiprobability $Q$ because of the product $\delta(b_1-\check{B}^c)\cdots \delta(b_n-\check{B}^c)$ with the same $\check{B}$, see also \ref{appx}. Let us add any noise to $b$, independent for every $\hat{B}$, $N=N_a(a)N_0(b_1)N_0(b_2)\cdots$. Let $(a_0,b_0+1)$ be the point where $N$ is maximal. Then $P(a_0,b_0,b_0,...)$ will be negative for a sufficiently long sequence because of multiplication of factors $N_0(b_0-1)$ which are smaller than $N_0(b_0+1)$ and the fact that $N_a(a_0)>N_a(a_0\pm 1)$. On the other hand, taking perfectly correlated noise $N_a(a)N_0(b_1)\delta(b_1-b_2)\cdots$ immediately  fails for the sequence $\hat{B}\hat{A}\hat{B}$.  In this case 
$P(b-1,a_0,b+1)$ will be negative if $N_a$ is maximal in $a_0$ because it has no help from $Q(\cdot,\cdot,b=+1)$. 

The presented example shows that our proposal fails ideal infinite cases. However, they can be excluded from our considerations for two reasons: (i) real imperfections of ideal few-level models and (ii) not every observable must have 
an exact counterpart in objective reality -- the actual observable is only indirectly related to the two-level space. This is also why repeated measurements in this case are good to test the noise model, which is rarely done.  For a moment, only long coherence experiments \cite{ions} may need such special care. Therefore we propose an approach based on moments and cumulants.
For $P$, $N$ and $Q$ we can define generating function $\Phi(k)=\langle\exp\: \rmi(k_aa+k_bb+k_cc+...)\rangle$ and 
moments $M_{pqr...}=\langle a^p b^rc^q\cdots\rangle$. Moments and cumulants can be obtained by differentiation of $\Phi$ and $\ln\Phi$ at $k=0$, respectively,
\begin{eqnarray}
&&M_{pqr...}=\frac{\partial^p}{\partial (\rmi k_a)^p}\frac{\partial^q}{\partial (\rmi k_b)^q}
\frac{\partial^r}{\partial (\rmi k_c)^r}\cdots\Phi|_{k=0}\nonumber\\
&&C_{pqr...}=\frac{\partial^p}{\partial (\rmi k_a)^p}\frac{\partial^q}{\partial (\rmi k_b)^q}
\frac{\partial^r}{\partial (\rmi k_c)^r}\cdots\ln\Phi|_{k=0}
\end{eqnarray}
In fact, $M_{pqr...}$ can be expressed by combinations of cumulants $C_{p'q'r'...}$ for $p'\leq p$, $q'\leq q$, $r'\leq r$, ... and viceversa, e.g $C_1=M_1$, $M_2=C_1^2+C_2$, $M_{11}=C_{10}C_{01}+C_{11}$ (note that $M_0=1$, $C_0=0$).  As generating functions factorize $\Phi_P=\Phi_N\Phi_Q$, cumulants from independent distributions simply add up, as in our case, $C_P=C_N+C_Q$. 
All moments with respect to $P$ can be expressed in terms of (finite combinations of) moments with respect to $N$ and $Q$.
This does not limit the total number of observables but only their number in a single average.
Apart from ideal models, quantum mechanics, by e.g. Keldysh formalism \cite{ctp} makes predictions rather in the form of moments (correlations) than full probability, hence this approach is more practical.
Moreover, since the present standard model of quantum mechanics must break down at extremely small distances and times (or high energies) it is plausible that it also cannot reliably describe very high moments.
In frequent situations of discrete outcomes, which might require full probability, one can 
also rewrite the problem in terms of moments, e.g. Bell test \cite{bb11}, where the condition $a=\pm 1$ is replaced by
$\langle(a^2-1)^2\rangle=0$.
Then the nonnegativity of $P$ follows from $\langle f(a,b,c,...)\rangle\geq 0$ for any real-valued positive polynomial $f\geq 0$,
e.g. $(a-b^2)^2$, $a^2b^4+a^4+b^2-3a^2b^2$ or $1+a^2b^2+b^2c^2+c^2a^2-4abc$. We shall demand nonnegativity of $P$ in this weaker sense, that only positive polynomials $f$ up to certain degree $K$ satisfy $\langle f\rangle\geq 0$, which is expressed by moments up to the same degree. Then is it enough to take $N$ as a Gaussian noise, which has only nonzero $C_2$ (variance). Sufficiently large $C_2$ will always make $\langle f\rangle$ positive. It is evident if we reconstruct first some $Q$ for discrete events \emph{near} origin, given the known moments.  Then we put events with small probability \emph{far} from the origin in all main directions, say  $Q(a\sim\pm\lambda\to\infty)\sim 1/\lambda^{K+1}$, and correct the quasiprobabilites $\sim 1/\lambda$ of the near events to restore original moments. A convolution of Gaussian $N$ with variance $\langle a^2\rangle\sim\lambda$ than makes $P$ positive, because it will be almost constant (and positive) around near events and elsewhere dominated by far (also positive) ones, from the fact that $\sum_{a'\sim\pm\lambda}\rme^{-(a-a')^2/\lambda}/\lambda^{K+1}\gg e^{-a^2/\lambda}$ for $|a|\gtrsim\lambda$. This demonstrates the positivity of $P$ in the weak sense. Note that the case (\ref{pnn}) will cause problems only if we let $K$ be unbounded.

\section{Relativistic invariance}

We ask if $P$, $N$ and $Q$ can be constructed without introducing any preferred frame, except when defined by the quantum state itself. Although not absolutely necessary, such condition could be desired.
As we shall see below, this is impossible if the observable is a vector field and barely possible for scalar fields.
Nevertheless, our construction outlined in Sec. \ref{secon} can be made but only in some preferred frame. 

If both $g$ and $f$ are invariant then all correlators with respect to $Q$ are also invariant, especially in the invariant vacuum state \cite{ab13}, so the only possible problem can be with $N$.
We denote Fourier transform $A(p)=\int dx A(x)\rme^{\rmi x\cdot p}$.
For any normal probability,
we have Cauchy-Schwarz type inequality
\begin{equation}
|\langle a(p_A)a(p_B)a(p_C)a(p_D)\rangle|^2\leq\langle|a(p_A)|^2\rangle\langle|a(p_B)a(p_C)a(p_D)|^2\rangle\label{cs1}
\end{equation}
with some regularization $\delta\to\delta_\epsilon$, i.e. $a(p)\to \int \rmd p' a(p')\delta_\epsilon(p-p')$.
Let us take a vector field (current) $a=j^\mu$ and the vacuum zero-temperature state, which is itself invariant \cite{ab13}. Then, for an arbitrary stationary and relativistically invariant probability, we have
$\langle j^\mu(p)j^\nu(q)\rangle=\delta(p+q)(p^\mu p^\nu\xi(p\cdot p)+g^{\mu\nu}\eta(p\cdot p))$
which must be positive definite. For spacelike $p$ it is positive only if $\eta=0$ (we could even use charge conservation to set $\xi=0$). Taking $p_A=(q,p,0,0)$ for $|q|<|p|$ we have then $\langle |j^3(p_A)|^2\rangle=0$ -- for the regularization we should replace $j^3$ by $(p_A\cdot p_A)j\cdot z -(j\cdot p_A)(p_A\cdot z)$ with $z=(0,0,0,1)$. However, the expression
$\langle j^3(p_A)j^3(p_B)j^3(p_C)j^3(-p_A-p_B-p_C)\rangle_Q$  usually does not vanish in the Markovian scheme ($f=0$), see an example in \ref{appb} if $g(p_A)\neq 0$ so the inequality (\ref{cs1}) is clearly violated.

In the non-Markovian case we can take the invariant generalization of $f$ derived in \cite{bbrb} which is $f(p)=i\mathrm{sgn}\:p^0 \theta(p\cdot p)$. However, then
fourth order correlations in vacuum still vanish. It follows from unitarity which implies vanishing
$\langle \check{A}^q\check{A}^q\check{A}^q\check{A}^q\rangle$.
Then $\langle \check{A}^c\check{A}^c\check{A}^c\check{A}^c\rangle=\langle\check{A}^+\check{A}^-\check{A}^-\check{A}^-\rangle+(+\leftrightarrow -)$ 
(right-hand side symmetrized). By Feynman rules nonvanishing crossing $+-$ must be timelike ($p\cdot p>0$)
\cite{ctp,ab13}.
Hence, our expression vanishes if all the Fourier arguments of $A$ are spacelike which is the case when all functions $f$ are zero.
We must take 6th order correlations, e.g. the Cauchy-Schwarz inequality
\begin{eqnarray}
&&|\langle a(p_A)a(p_B)a(p_C)a(-p_A)a(-p_B)a(-p_C)\rangle|^2\leq\nonumber\\
&&\langle|a(p_A)|^2\rangle\langle|a(p_B)a(p_C)a(-p_A)a(-p_B)a(-p_C)|^2\rangle\label{cs2}
\end{eqnarray}
is violated by nonvanishing left hand side (example in \ref{appb}). 
The important consequence is that we cannot construct any invariant objective realism, using (\ref{conv}) and (\ref{avgabs}) for vector fields.  This conclusion is  completely general if only we can somehow argue that the left hand side is nonzero.  It does not prevent from construction of non-invariant realism, see also next section, but puts relativity at stake, it is possible that not only objective realism fails to be invariant but the quantum theory itself
at extremely high energies.

For scalar fields $a=\phi$ we can formally make the autocorrelation $\langle|\phi(p)|^2\rangle$ positive for all $p$.
However, from the \emph{correspondence principle} follows that zero-frequency (long-time) quantum equilibrium correlations satisfy classical fluctuation-dissipation theorem \cite{fdt}, which makes them vanish in vacuum (zero-temperature). This fact  follows directly from closed time path formalism \cite{ctp,ab13} and does not rely at all on relativity. On the other hand, correlations can be invariant only at zero temperature vacuum. Then all vacuum zero-frequency correlations must vanish and, from invariance,   $\langle |\phi(p)|^2\rangle$ must vanish for all spacelike $p$ so the inequality (\ref{cs1}) or (\ref{cs2}) will be violated by a scalar analogue of the previous example. This shows that autocorrelation for $N$ at spacelike $p$ must violate correspondence principle. Another attempt to restore invariance would be a classical filter permitting only timelike momenta, i.e. $g(p)=0$ for spacelike $p$. Unfortunately it would violate relativistic causality, as for invariant $g$, for spacelike $x$ we get $g(l=\sqrt{-x\cdot x})=\int ds K_1(sl)g(s=\sqrt{p\cdot p})s^2/2l\pi^3$ (with Bessel $K$) \cite{fourl} which is always somewhere nonzero.
It follows rigorously from nonvanishing Fourier transform of $K_1(e^q)$ which can be expressed by Gamma functions,
see e.g. \cite{k1f}. We cannot also take $f(p)=\pm \rmi\theta(-p\cdot p)$ as it gives complex $f(x)$. The invariance and correspondence principle at zero temperature will be hence broken in our construction of objective reality, at least spontaneously. Once we accept this, it is easy to find an analogy with a particular Gaussian measurement scheme (\ref{appc}), apparently invariant \cite{csl}, but supplemented by a heat sink in a preferred frame to become stationary. Only formalisms beyond standard correspondence principle or quantum field theory may allow relativistically invariant realism \cite{beding}.

\section{Noise}

The noise $N$ can of course be partly from some coexistent system or universe, but never all, due to the violation of invariance and correspondence principle.
Once we learned that the noise cannot be invariant, we can define it in a preferred frame. We shall assume that $N$ is white Gaussian and zero-centered with the correlations
\begin{eqnarray}
&&\langle \phi(x)\phi(x')\rangle_N=n\delta(x-x'),\\
&&\nonumber\langle j^\mu(x)j^\nu(x')\rangle_N=(2g^{\mu 0}g^{\nu 0}-g^{\mu\nu})n_j\delta(x-x')
\end{eqnarray}
with positive $n$, $n_j$. This is not a quantum amplitude but classical correlation function.
The noise will make $P$ positive for a finite number of moments, depending on $n$, as we have shown in general in Sec. \ref{secq}. The exact relation between $n$ and the order of moments needs to be found for concrete quantum field theories.
Noise correlations obeying these requirements are analogous as in the general quantum measurement formalism, based on
Kraus operators  \cite{kraus}, see appendix \ref{appc}, which is a straightforward extension of the analysis of \cite{bbrb}.
The value of $n$ can be only estimated from the fact that the noise is invisible macroscopically. Taking electric 
charge, suppose that we can observe a capacitor in a $1$mm box in a tenth second, which contains about $10^{20}$ atoms 
(electrons and protons, to simplify). This gives an estimate $n_j\sim 3\cdot 10^{41}e^2/$m$^4$. On the other hand at the 
level of single atoms, the relevant scales are: length $10^{-10}$m, time (in meters) $10^{-7}$m, which gives the noise 
constant to compare $\sim 10^{37}e^2/$m$^4$. Therefore the inherent quantum noise, and also all higher correlations, 
which rely on the same scales, can be much smaller than the external noise. If we make the noise non-white, 
frequency-dependent, the estimate can be pushed even much higher. In long coherence setups \cite{ions}, we can add a resonant 
noise. That noise is also decoupled from all dynamical quantum evolution, so there is in principle no way to measure 
it, except estimation based on personal perception. It can be even much larger than our estimate and only common sense 
tell us when it becomes unrealistic. 

\section{Conclusion}
 We have constructed objective quantum reality developed on weak, noninvasive measurements and satisfying linearity and causality but not relativistic invariance. The reality is represented by a nonnegative probability measure for observations and its cumulants/moments, a convolution of quantum quasiprobability and external noise. The latter cannot be relativistically invariant and must violate correspondence principle. These two principles, valid so far both in quantum and classical physics, but impossible in quantum objective realism, are therefore put at stake and further theoretical and experimental studies are necessary to check their general validity. A reasonable value of the noise ensures that the net probability will be positive, while remaining macroscopically unobservable. An open question remains whether one can make this objective reality \emph{local}, which however, needs further generalization, including freedom of choice.

\section*{Acknowledgements}
I thank W. Belzig, P. Chankowski and P. Pearle for discussions, hints and inspirations that helped me to complete this work.

\appendix
\section{Connection between quasiprobability $Q$ and weak measurement}
\label{appa}

We shall derive the form (\ref{avgabs}) and (\ref{avg2}) by taking noninvasive limit
in the general quantum measurement formalism, based on
Kraus operators $\hat{K}$ \cite{kraus}.
The probability distribution
of the measurement results is $\rho=\langle\check{K}\rangle$ for
$\check{K}\hat{X}=\hat{K}\hat{X}\hat{K}^\dag$, where the only
condition on $\hat{K}$ is that the outcome probability is normalized
regardless of the input state $\hat{\rho}$. Here we need $\hat{K}$ to
be spacetime-dependent. In general, we assume that $\hat{K}[\hat{A},a]$ is
a functional of the whole spacetime history of observables $\hat A(x)$ and
outcomes $a(x)$. We shall assume that the functional $\hat{K}$ is
stationary so it depends only on relative spacetime arguments.

The essential step to satisfy Eq.~(\ref{avgabs}) is to take the limit
$\hat{K}\sim \hat{1}$
which corresponds to a noninvasive measurement.
This can be obtained from an arbitrary initial POVM by rescaling
$\hat{K}[\hat{A},a]\to \hat{K}_\eta =C(\eta)\hat{K}[\eta\hat{A},\eta
  a]$ with $\eta\to 0$, which defines
$\rho_\eta=\langle\check{K}_\eta\rangle$. Here $C(\eta)$ is a
normalization factor.

The desired correlation function (\ref{avgabs}) can be derived by the
following limiting procedure for an almost general POVM,
\begin{equation}
\langle a_1(x_1)\cdots a_n(x_n)\rangle_Q=
\lim_{\eta\to 0}\langle a_1(x_1)\cdots a_n(x_n)\rangle_\eta\:,
\end{equation}
where the average on the right-hand side is with respect to
$\rho_\eta$.  We assume the absence of internal correlations between
different detectors, namely $\hat{K}[\hat{A},a]=\mathcal
T\prod_j\hat{K}[\hat{A}_j,a_j]$, where $\mathcal T$ applies to the
time arguments of $\hat{A}$.

Expanding
$\hat{K}[\hat A,a]/k[a]=1+\int \mathrm{d}x' F[a,x']\hat A(x') +
\mathcal O(\hat A^{2})$,
we find, up to $\mathcal O(\hat{A}^{2})$,
\begin{equation}
\check{K}/|k[a]|^2\simeq 1
+\int \rmd x'\;\left(2\mathrm{Re}F\check{A}^c(x')
-\mathrm{Im}F\check{A}^q(x')\right)\:.
\label{main_result}
\end{equation}
Here, $|k[a]|^2$ is a functional probability of time-resolved outcomes independent of
the properties of the system which represents the detection error.
As we want the measurement to be noninvasive to lowest order, we
impose the condition that $\int F[a,x']|k[a]|^2\mathrm{D}a$ vanishes;
$\mathrm{D}a$ is the functional measure.  Our conditions imply that
$\int 2a(x)\mathrm{Re}F[a,x']|k[a]|^2\mathrm{D}a=g(x-x')$, and
we get $f(x-x')=-\int 2a(x)\mathrm{Im}
F[a,x']|k[a]|^2\mathrm{D}a$.  Thus, the most general weak Kraus
operator takes the form given in Eq.~(\ref{main_result}).

\section{Discrete Wigner function}
\label{appx}

Here we calculate in detail the quasiprobability in Table \ref{tab1}, for first $\hat{A}$ then $\hat{B}$ represented
in the basis $|+\rangle,|-\rangle$ by
\begin{equation}
\hat{A}=\left(\begin{array}{rr}
0&1\\
1&0\end{array}\right),\:
\hat{B}=\left(\begin{array}{rr}
1&0\\
0&-1\end{array}\right)
\end{equation}
with eigenvalues $+1,-1$.
For simplicity  $f=0$ (Markovian case) we have $\check{A}=\check{A}^c$ and both $\check{A}$ and $\check{B}$ can be represented as $4\times 4$ matrices in the basis $|+\rangle\langle+|,|+\rangle\langle -|,|-\rangle\langle +|,|-\rangle\langle -|$,
\begin{equation}
\check{A}=\frac{1}{2}\left(\begin{array}{rrrr}
0&1&1&0\\
1&0&0&1\\
1&0&0&1\\
0&1&1&0\end{array}\right),\:
\check{B}=\left(\begin{array}{rrrr}
1&0&0&0\\
0&0&0&0\\
0&0&0&0\\
0&0&0&-1\end{array}\right).
\end{equation}
Both matrices have eigenvalues $1,0,0,-1$, and the corresponding eigenstates
\begin{eqnarray}
&&\frac{1}{2}\left(\begin{array}{r}1\\1\\1\\1\end{array}\right),
\frac{1}{2}\left(\begin{array}{r}1\\1\\-1\\-1\end{array}\right),
\frac{1}{2}\left(\begin{array}{r}1\\-1\\1\\-1\end{array}\right),
\frac{1}{2}\left(\begin{array}{r}1\\-1\\-1\\1\end{array}\right);\nonumber\\
&&
\left(\begin{array}{r}1\\0\\0\\0\end{array}\right),
\left(\begin{array}{r}0\\1\\0\\0\end{array}\right),
\left(\begin{array}{r}0\\0\\1\\0\end{array}\right),
\left(\begin{array}{r}0\\0\\0\\1\end{array}\right).
\label{eqab}
\end{eqnarray}
Our initial state is $\hat{\rho}=|+\rangle\langle +|=(1,0,0,0)^T$. Applying (\ref{qqq}), the objects $\delta(b-\hat{B})$ and  $\delta(a-\check{B})$ essentially mean replacing $\hat{B}$  and $\check{B}$ by their eigenvalues,
\begin{equation}
\left(\begin{array}{cc}
\delta(b-1)&0\\
0&\delta(b+1)
\end{array}\right),\:
\left(\begin{array}{cccc}
\delta(b-1)&0&0&0\\
0&\delta(b)&0&0\\
0&0&\delta(b)&0\\
0&0&0&\delta(b+1)
\end{array}\right)
\end{equation}
and essentially the same for $a$ except different eigenbasis. The calculation of $Q(a,b)=\mathrm{Tr}\delta(b-\hat{B})\delta(a-\check{A})\hat{\rho}$  (we can replace $\check{B}$ by $\hat{B}$) is conducted as follows.
The find projection of $\hat{\rho}$ onto eigenstates of $\check{A}$ obtaining each time $1/2$ and values $a=1,0,0,-1$.
From the trace and diagonal form of $\hat{B}$, we extract only the first and last entry, namely
$(1,1)$, $(1,-1)$, $(1,-1)$, $(1,1)$, respectively, with the final weight $1/4$. It yields all final events and their probabilities. Note that $b=0$ cannot appear but there are two cases $a=0$, giving finally $\pm 1/2$ as shown in Table \ref{tab1}.

\section{Nonzero correlations with spacelike field}
\label{appb}

\begin{figure}
\includegraphics[scale=.5]{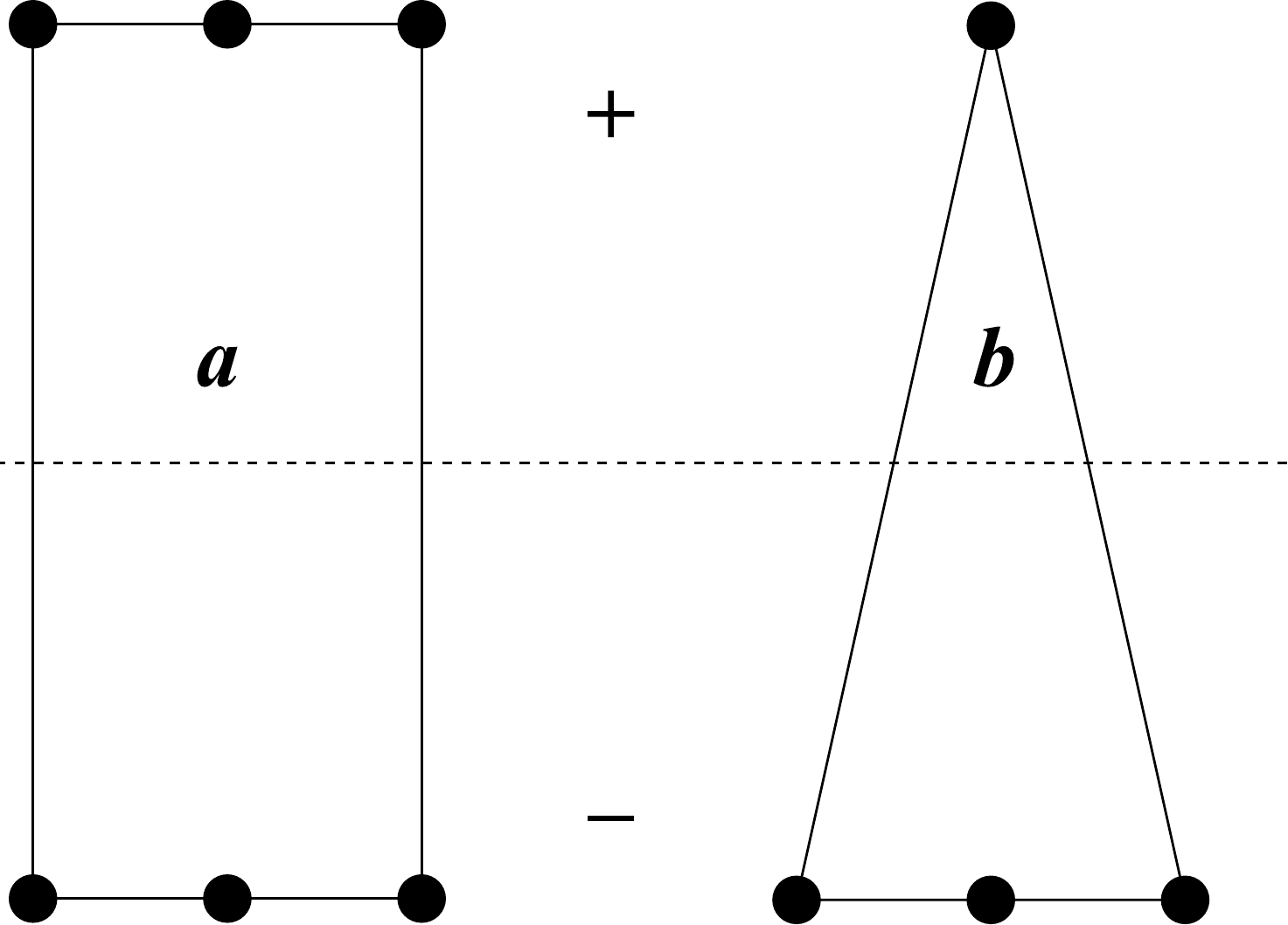}
\caption{Closed time path Feynman diagrams with electron propagators as lines, measured currents as black points, and $+/-$ are forward/backward (upper/lower) fields or equivalent operators.
(a) sixth order current correlation (\ref{333}) used in (\ref{cs2}) (b) fourth order correlation
used in (\ref{cs1})}\label{diagg}
\end{figure}

Sixth order current correlations, by unitarity rule $\langle \check{A}^q\check{A}^q\cdots\check{A}^q\rangle=0$ and closed time path formalism \cite{ctp,ab13} can be expressed as
$\langle \check{j}^c\check{j}^c\check{j}^c\check{j}^c\check{j}^c\check{j}^c\rangle=\langle\check{j}^+\check{j}^-\check{j}^-\check{j}^-\check{j}^-\check{j}^-\rangle+\langle\check{j}^+\check{j}^+\check{j}^+\check{j}^-\check{j}^-\check{j}^-\rangle+(+\leftrightarrow -)$. The nonvanishing term has the form $\langle\check{j}^+\check{j}^+\check{j}^+\check{j}^-\check{j}^-\check{j}^-\rangle$.
As example, let us take $\langle |j^0(p_A)j^0(p_B)j^0(p_C)|^2\rangle$ where
$p_X=(2q_0/3,q_\perp\cos\phi_X,q_\perp\sin\phi_X,0)$ with $q_0\gtrsim m$ (electron mass), $\phi_{A,B,C}=0,2\pi/3,4\pi/3$ so that spatial vectors point symmetrically in the plane while $q_0$ is just above decay threshold and $q_\perp>q_0$ so that $p\cdot p<0$.
The lowest order contribution has the form (depicted graphically by Feynman diagrams \cite{peskin} in Fig. \ref{diagg}a)
\begin{eqnarray}
&&\int \rmd k \mathrm{Tr}\frac{\gamma\cdot (k +q_0-p_C)+m}
{(k + q_0-p_C)^2-m^2}\gamma^0\frac{\gamma\cdot(k-q_0+p_A)+m}
{(k-q_0+p_A)^2-m^2}\gamma^0\times\nonumber\\
&&(\gamma\cdot(k-q_0)+m)
\gamma^0\frac{\gamma\cdot( k -q_0+ p_C)+m}{(k -q_0+p_C)^2-m^2}\gamma^0\frac{\gamma\cdot(k+q_0-p_A)+m}{(k+q_0-p_A)^2-m^2}
\nonumber\\
&&\gamma^0(\gamma\cdot( k+q_0)+m)\gamma^0\delta^+_m(k-q_0)\delta^-_m(k+q_0)+\mathrm{sym}\label{333}
\end{eqnarray}
where $q_0=(q^0,0,0,0)$, $\gamma^\mu$ are Dirac $4\times 4$ matrices such that $\{\gamma^\mu,\gamma^\mu\}=2g^{\mu\nu}$, and $(X)^2=X\cdot X$.
In the expression 
one must perform full symmetrization of two separate sets $ABC$ for the left and right branch. The $\delta_\pm$ correspond to transitions $+-$ and here just make $k$ almost irrelevant, $k\sim 0$.
In this case all denominators have almost the same strictly negative value. Evaluation of the numerator trace and symmetrization gives for $q_\perp\gg q^0$ the leading term $q_\perp^4$ with some nonvanishing dimensionless factor so 
(\ref{333}) is nonzero. 
The  fourth order correlator appearing in (\ref{cs1}),
$\langle j^3(p_A)j^3(p_B)j^3(p_C)j^3(-p_A-p_B-p_C)\rangle_Q$ (depicted in Fig. \ref{diagg}b) does not vanish in the Markovian scheme ($f=0$) analogously as (\ref{333}). Unitarity 
$\langle \check{A}^q\check{A}^q\check{A}^q\check{A}^q\rangle=0$ helps to simplify the relevant expression $\langle \check{A}^c\check{A}^c\check{A}^c\check{A}^c\rangle=\langle\check{A}^+\check{A}^-\check{A}^-\check{A}^-\rangle+(+\leftrightarrow -)$ (right-hand side symmetrized).

\section{Gaussian example of Kraus operators}
\label{appc}

An example of a POVM leading to relativistically invariant $Q$ in the weak limit is
based on the Gaussian detector prepared in the initial state (wavefunction) $\phi(X)\propto\exp(-X^2)$, where $X$ is not a fourvector $x$ but simple one-dimensional measurable position-like parameter,
interacting with the system by the time-dependent Hamiltonian density (in the interaction picture)
$\hat{\mathcal H}_I(x')=(\delta(x-x')\hat{P}+2f(x-x')\hat{X})\hat{A}(x')$. The momentum-like quantity $\hat{P}$ makes the shift $\hat{X}\to \hat{X}-\hat{A}(x)$.
For the measurement of $a(x)=X$ we get
the Gaussian Kraus operator
\begin{eqnarray}
&&\hat{K}[\hat{A},a]\propto\label{keq}
\mathcal T \exp\left[-(\hat{A}(x)-a(x))^2
-\vphantom{\int}\right.\\
&&\nonumber\left.
\int \rmd x'\;2if(x-x')(a(x)-\hat{A}(x)\theta(x^0-x^{\prime 0}))\hat{A}(x')\right]
\end{eqnarray}
Here, the first term in the exponent is the Markovian part, while the
second term describes the non-Markovian measurement process including
a fixed but arbitrary real function $f(x)$, characterizing the memory
effect, as it makes the outcome depend on distant observables. The Heaviside function $\theta$ follows from the fact that $P$ shifts the phase for $x^{\prime 0}<x^0$ and ensures the
normalization of the Kraus operator.
By comparing with (\ref{main_result}), we get $|k[a]|^2=\sqrt{2/\pi}\rme^{-2a^2}$
and $F[a,x']=2a(x)(\delta(x-x')-if(x-x'))$, which in this special case are just usual functions.
Following the standard procedure
we find the Kraus superoperator in the form
\begin{eqnarray}
&&\check{K}[\hat{A},a]\propto
\mathcal T \exp\left[-2(\check{A}^c(x)-a(x))^2+
\frac{(\check{A}^q(x))^2}{2}
\vphantom\int+\right.\\
&&\left.\int \mathrm{d}x'\;2f(x-x')(a(x)\check{A}^q(x')-
\theta(x^0-x^{\prime 0})(\check{A}^c(x)\check{A}^q(x')+
\check{A}^q(x)\check{A}^c(x')))\right]\nonumber
\end{eqnarray}

To prove the normalization, $\int da
\langle\check{K}\rangle=1$, we perform the Gaussian integral
over $a$ (time order is no problem if kept up throughout the
calculation) and get
\begin{eqnarray}
&&\int \rmd a\;\check{K}=\mathcal T
\exp\left[(\check{A}^q(x))^2/2+\vphantom\int\right.\\
&&
\int \rmd x' \theta(x^{\prime 0}-x^0)
2f(x-x')\check{A}^q(x')\check{A}^c(x)\mathrm{d}x'\nonumber\\
&&-\theta(x^0-x^{\prime 0})2f(x-x')\check{A}^q(x)\check{A}^c(x')\rmd x'\nonumber\\
&&\left.
+ f(x-x')f(x-x'')\check{A}^q(x')
\check{A}^q(x'')\rmd x'\rmd x''/2\vphantom\int\right],\nonumber
\end{eqnarray}
where we have ordered properly $\check{A}^q(x')$ and $\check{A}^c(x)$.
In the power expansion, omitting the identity term, the leftmost
superoperator is always $\check{A}^q$. Since
$\mathrm{Tr}\check{A}^q\cdots=0$ we obtain $\int
\mathrm{d}a\;\langle\check{K}[\hat{A},a]\rangle=1$ or $\int
\mathrm{d}a\; \hat{K}^\dag\hat{K}=\hat{1}$.  In general, we define
$\hat{K}[\hat{A},a]$ for $n$ measurements as
$\hat{K}[\hat{A},a]=\mathcal T\prod_j \hat{K}[\hat{A}_j,a_j]$, taking $\hat{H}_I=\sum_j\hat{H}_{j,I}$. 
To get a weak measurement, we substitute $\hat{K}$ by $\hat{K}_\eta$ which is obtained by
replacing $\hat{H}_I\to\eta\hat{H}_I$ and measuring $a(x)=\eta X$.
Note
that putting $\hat{A}=0$ gives Gaussian white noise $\rho\propto
\rme^{-2a^2}$, which leads to large detection noise in the weak
limit, $\rho_\eta\propto e^{-2\eta^2 a^2}$, that has to be
subtracted/deconvoluted from the experimental data. The scheme is apparently relativistically invariant just as shown in \cite{csl} but for any finite $\eta$ the disturbance heats the system constantly making it impossible to reach a stationary state. The heat must be transferred to a sink in a preferred frame.

\end{document}